\documentclass[12pt,oneside,onecolumn]{IEEEtran}
\usepackage{subfigure,graphicx,dsfont,amssymb,amsmath,cite,calc,tabulary,setspace}
\usepackage{xcolor}

\begin{document}
	\title{Interference Alignment in MIMO Interference Channels using SDP Relaxation}
	\author{Siavash~Mollaebrahim$^*$, Pouya~Ghari$^\dagger$, \textit{Student Member}, \textit{IEEE}, Muhammad Sadegh~Fazel$^*$, Muhammad Ali Imran, \textit{Senior Member}, \textit{IEEE}
		\thanks{S.Mollaebrahim and M.Fazel are with Electrical Engineering Department, Isfahan University of Technology (IUT), Isfahan, Iran, e-mail: siavashmol@gmail.com, fazel@cc.iut.ac.ir
			
			P.Ghari is with the Electrical Engineering Department, University of Tehran (UT), Tehran, Iran, e-mail: p.m.ghari@ut.ac.ir
			
			M.A. Imran is with the School of Engineering, University of Glasgow, UK, e-mail: Muhammad.Imran@glasgow.ac.uk }}
\maketitle
\begin{abstract}
Nowadays, providing higher data rate is a momentous goal for wireless communications systems. Interference is one of the important obstacles to reach this purpose. Interference alignment is a management technique that align interference from other transmitters in the least possible dimension subspace at each receiver and as a result, provide the remaining dimensions for free interference signal. An uncoordinated interference is an example of interference which cannot be aligned coordinately with interference from coordinated part and consequently, the performance of interference alignment approaches is degraded. In this paper, we propose two rank minimization methods to enhance the performance of interference alignment in the presence of uncoordinated interference sources. Firstly, a new objective function is chosen then, a new class of convex relaxation is proposed with respect to the uncoordinated interference which leads to decrease the optimal value of our optimization problem. Moreover, we use schatten-p-norm as surrogate of rank function and we implement iteratively reweighted algorithm to solve optimization problem. In addition, we apply our proposed methods to mitigate interference in relay-aided MIMO interference channel, and propose a weighted-sum method to improve the performance of interference alignment in the amplify-and-forward relay-aided MIMO system based on the rank minimization approach. Finally, our simulation results show that our proposed methods can obtain considerably higher multiplexing gain and sum rate than other approaches in the interference alignment framework and the performance of interference alignment is improved. 
\end{abstract}
\section{Introduction}
\IEEEPARstart{I}{nterference} is an important problem in wireless networks and may cause severe limitations in transmitting information; therefore, it is essential to develop communication schemes in order to manage interference. Recently, several methods have been proposed to deal with interference. Most of these methods are based on two basic approaches: either accounting other users’ signal as a noise or orthogonalizing communication links. Using these two types of approaches for a $k$-user MIMO system, a portion $1/k$ of whole resources are available to each user. The main drawback of these approaches is that when the number of users is increased, available resources to each user are decreased significantly \cite{bresler2014feasibility}. One of the most effective techniques to manage interference is interference alignment (IA). The key idea of the IA is to align the whole interference signals at each receiver, and to make interference and desired signal subspaces linearly independent from each other. In the high signal-to-noise ratio (SNR) region, degrees of freedom (DOF) also known as multiplexing gain is the first order approximation of the sum capacity. Thus, the sum-capacity of interference channels can be characterized by DOF (or multiplexing gain) \cite{cadambe2008interference}. It has been shown that by using IA in the $K$-user MIMO system where each user has $N$ antennas, the DOF of $KN/2$ can be obtained \cite{Gou2010}. However, this result
is highly dependent on the channel extensions, and practically the channel extensions is not usually available. Moreover, it
can increase the complexity of implementation \cite{chen2014interference}. The DOF can be interpreted as free signaling dimensions \cite{papailiopoulos2012interference}. Practically, to obtain predicted DOF by IA, precoding matrix and receiver filter should be designed. Indeed, designing precoding and receiver filter matrices is simpler to implement than asymptotic interference alignment schemes such as \cite{cadambe2008interference} which needs to decomposition of multi-antenna nodes and infinite symbol extensions \cite{Gokul2015}.

Rank constrained rank minimization (RCRM) has been proposed in order to design precoding and receiver matrices for IA \cite{papailiopoulos2012interference} . Authors of \cite{papailiopoulos2012interference}  show that, minimizing rank of interference matrix can decrease dimension of interference subspaces in order to obtain DOF as large as possible. Due to the fact that the rank minimization problem is NP-hard and non-convex, the nuclear norm heuristic method has been employed \cite{papailiopoulos2012interference}. In this method, sum of magnitudes of singular values is minimized as a convex approximation of the rank function whereas the rank function is the number of non-zero singular values. Another convex approximation has been proposed in \cite{du2013reweighted} based on log function.    

The heterogeneous networks (HetNets) consist of macro cell networks and various small cells, these networks can meet the explosively increasing traffic demand in wireless networks. Moreover, by using small cells such as pico-cells and femto-cells the indoor coverage of cellular networks is noticeably increased \cite{Liu2015}. However, using small cells can cause severe interference which is a serious obstacle to reach advantages of small cells in HetNets; therefore, interference management is an important issue.

In this paper, we consider one of the most common types of interference in wireless networks which is named as uncoordinated interference. An interference source that is not coordinated by network can cause such interference \cite{peters2011cooperative}. For example, in the heterogeneous pico-cell networks, interference caused by femtos and home base stations is considered as uncoordinated interference, and cannot be fully aligned. In the heterogeneous network, the information about the uncoordinated interference can be obtained. The presence of uncoordinated interference degrades the performance of the coordinated part and cannot be ignored \cite{damnjanovic2011survey}.
  
 Furthermore, interference is an important issue for MIMO relay-aided networks. Infact, interference impacts the received signal at the relay and also the received signal at the destination; consequently, relay based communication is highly sensitive to interference. When a number of pairs of users communicate with each other via help of MIMO relays, the design of relay processing matrices is complicated because each pair of users suffers from interference produced by other users. In the high SNR regime, the destructive effect of interference is become more serious; therefore, interference should be handled in MIMO relay networks \cite{zhang2015}. Amplify-and-forward (AF) relaying is more suitable than decode-and-forward (DF) relaying
 to use in the MIMO communication system (especially in heterogeneous networks), due to the fact that AF relay
 has a lower complexity and shorter delay compare to DF relaying \cite{berger2009}. Despite the remarkable development of IA in the single-hop interference channel, IA is not notably advanced in the relay-aided interference channel. Unfortunately, previous  algorithms for single-hop interference alignment are not applicable for relay-aided IA because power constraints of relays are dependent on the precoding matrices at transmitters  and also the processing matrices at relays. Moreover, there are a few algorithms for relay-aided interference alignment. Therefore, proposing new algorithms to offer better solutions is important. In \cite{Troung2013}, authors have proposed two approaches for AF relay-aided IA. In the first approach, the sum power of interference at receiver is minimized. In the second one, minimizing the weighted sum mean square errors (WMSE) is the aim of the algorithm. Both of them have been proposed based on approaches existed in the single-hop MIMO interference channel \cite{shi2011}, \cite{peters2009interference}. 
 
 In this paper, we propose two new rank minimization methods in IA framework to enhance the performance of IA, especially when the uncoordinated interference sources exist in the MIMO interference system. In addition, a new class of convex constraint is proposed which expands the feasibility set of the optimization problem. This relaxation considers the possible solutions that are overlooked by existing RCRM-based IA algorithms. For analysis we obtain the dual problem of IA. The analysis shows that the proposed constraint can reduce the optimal value of the optimization problem (which leads to higher DOF). Besides, we generalize our proposed methods to AF relay-aided MIMO system. To turn power constraints of the optimization problem into a convex term a semi-definite programming (SDP) relaxation is used, then in order to solve bi-objective optimization problem and find pareto optimal point a weighted sum method is proposed.  Simulation results confirm that our proposed methods can achieve higher DOF and sum rate compare to other approaches. 
 
  The organization of the remainder of this paper is as follows. Section II describes the problem formulation. Section III presents our proposed methods. Section IV evaluates numerically the proposed algorithms. Section V concludes this paper. Sections VI-X represents Appendices.
  
\emph{Notation}: Bold uppercase letters such as $\boldsymbol{A}$ denote matrices, bold lowercase letters such as $\boldsymbol{a}$ denote column vectors. The expectation operator is denoted by $E\left[ . \right]$. Hermitian transpose of matrix $\boldsymbol{X}$ and $i$-th largest singular value are $\mathop{\boldsymbol{X}}^{H}$ and $\mathop{\delta }_{i}(\boldsymbol{X})$, respectively. Trace of matrix $\boldsymbol{X}$ is $tr(\boldsymbol{X})$. $\mathop{\small\boldsymbol{1}}_{d}$  and  $\mathop{\boldsymbol{I}}_{d}$ are all-one and identity d by d matrices, respectively. Nuclear norm of matrix $\small\boldsymbol{X}$ can be stated as $\mathop{\left\| \boldsymbol{X} \right\|}_{*}=\sum\nolimits_{i=1}^{rank(\boldsymbol{X})}{\mathop{\delta }_{i}(\boldsymbol{X})}$. $\boldsymbol{X}(i,j)$ determines the element in the $i$-th row and $j$-th coloumn of $\boldsymbol{X}$.
\section{Problem Formulation}
\subsection{System Model for MIMO Interference Channel}
 We consider a case in which $K$-user interference wireless system contains $K$ transmitters and $K$ receivers. The number of antennas for each receiver and transmitter is denoted by $\mathop{M}_{r}$ and $\mathop{M}_{t}$, respectively. Transmitters are assumed to be synchronized, and each user transmits the symbol vector $\mathop{\boldsymbol{x}}_{k}\in \mathop{\mathbb{C}}^{d\times1}(k=1,...,K)$ to its associated receiver. This $K$-user MIMO system can be represented by ${{({{M}_{t}}\times{{M}_{r}},d)}^{K}}$ \cite{yetis2010feasibility}. In addition, $X$ sources of uncoordinated interference exist in the system, and $\mathop{\boldsymbol{y}}_{k}\in \mathop{\mathbb{C}}^{{M}_{r}\times1}$ is considered as the received signal in the $k$-th receiver. After linear processing, the output can be expressed as follows:
\begin{align}
	\boldsymbol{U}_{k}^{H}\boldsymbol{y}_{k}=\boldsymbol{U}_{k}^{H}\boldsymbol{H}_{k,k}\boldsymbol{V}_{k}\boldsymbol{x}_{k}+\small\boldsymbol{U}_{k}^{H}\sum\limits_{l=1,l\ne k}^{K}\boldsymbol{H}_{k,l}\boldsymbol{V}_{l}\boldsymbol{x}_{l}+\nonumber\\\boldsymbol{U}_{k}^{H}\sum\limits_{f=1}^{X}\boldsymbol{C}_{k,f}\boldsymbol{F}_{f}\boldsymbol{q}_{f}+\boldsymbol{U}_{k}^{H}\boldsymbol{w}_{k}
\end{align}
where $\mathop{\boldsymbol{H}}_{k,l}\in \mathop{\mathbb{C}}^{{M}_{r}\times{M}_{t}}$ represents the flat fading channel between $k$-th receiver and $l$-th transmitter. Linear receive filter and precoding matrices are denoted by $\mathop{\boldsymbol{U}_{k}}\in \mathop{\mathbb{C}}^{{M}_{r}\times{d}}$ and $\mathop{\boldsymbol{V}}_{k}\in \mathop{\mathbb{C}}^{{M}_{t}\times{d}}$, respectively. $\mathop{\boldsymbol{C}}_{k,f}\in \mathop{\mathbb{C}}^{{M}_{r}\times{M}_{f}}$ is the channel between the $k$-th receiver and the $f$-th uncoordinated source. The number of antennas of the $f$-th uncoordinated source is denoted by ${M}_{f}$. $\mathop{\boldsymbol{F}}_{f}\in \mathop{\mathbb{C}}^{{M}_{f}\times{d}_{f}}$ and $\mathop{\boldsymbol{q}}_{f}\in \mathop{\mathbb{C}}^{{d}_{f}\times{1}}$ represent the precoding matrix and symbol vector of the $f$-th uncoordinated source, respectively. Furthermore, ${d}_{f}$ denotes the length of data stream of the  $f$-th uncoordinated source. Finally, $\boldsymbol{w}_{k}$ is the zero-mean complex additive white Gaussian noise with covariance $\sigma _{k}^{2}{\boldsymbol{I}_{_{M{}_{r}}}}$. We assume that $E[{{\left\| {\boldsymbol{V}_{k}}{\boldsymbol{x}_{k}} \right\|}^{2}}]={{P}}$, $E[{{\left\| {\boldsymbol{F}_{f}}{\boldsymbol{q}_{f}} \right\|}^{2}}]={{P}_{f}}$ where $\mathop{P}$ is the transmit power of each user and $\mathop{P}_{f}$ is the transmit power of each uncoordinated source.

The required conditions for perfect IA can be stated as \cite{yetis2010feasibility}:
\begin{align}
	\boldsymbol{U}_{k}^{H}\boldsymbol{{H}}_{k,l}\boldsymbol{V}_{k}=0\\
	rank(\boldsymbol{U}_{k}^{H}\boldsymbol{H}_{k,k}\boldsymbol{V}_{k})=d
\end{align}
The system ${{({{M}_{t}}\times{{M}_{r}},d)}^{K}}$ is proper if $\frac{{M}_{r}+{M}_{t}}{K+1}\ge d\ $. Obviously if this condition is not satisfied, the system is improper \cite{yetis2010feasibility}. The authors of \cite{yetis2010feasibility} have shown that this condition is necessary but not sufficient in a general case. In addition, in the improper system, perfect IA cannot be admitted. In the later work on feasibility of IA, it was shown that when Mt = Mr the necessary condition is also sufficient \cite{cadambe2008interference}. Finaly, in \cite{ruan2013feasibility} the authors have shown that the necessary condition can be sufficient in more general case even though $Mt\ne{Mr}$. In improper system the rank of interference is not zero and the goal is to find out linear precoding and linear receive filter matrices that can minimize the rank of interference. Since there is uncoordinated interference in our scenario, the system is improper and perfect IA is not feasible \cite{ruan2013feasibility}. In this paper, we try to minimize the rank of interference by RCRM framework.

We define the signal and interference matrices $({\boldsymbol{S}_{k}},{\boldsymbol{J}_{k}})$ for all  $k= 1,...,K$ in the presence of the uncoordinated interference. 
\begin{equation}
	\boldsymbol{S}_{k}\triangleq {\boldsymbol{U}}_{k}^{H}\boldsymbol{H}_{k,k}\boldsymbol{V}_{k}
\end{equation}
\begin{equation}
	{{\boldsymbol{J}}_{k}}\triangleq\boldsymbol{U}_{k}^{H}\left[ \left\{ {{\boldsymbol{H}}_{k,l}}{{\boldsymbol{V}}_{l}} \right\}_{l=1,l\ne k}^{K}...\left\{ {{\boldsymbol{C}}_{k,f}}{{\boldsymbol{F}}_{f}} \right\}_{f=1}^{X} \right]
\end{equation} 
where ${{\boldsymbol{J}}_{k}}\in {{\boldsymbol{C}}^{d\times[(K-1)d+{{D}_{f}}]}}$. We consider ${D}_{f}=\sum\limits_{f=1}^{X}{{d}_{f}}$. In addition, the multiplexing gain of user $k$ can be expressed in terms of rank function:
${{DOF}_{k}}\triangleq rank({\boldsymbol{S}_{k}})-rank({\boldsymbol{J}_{k}})$\cite{papailiopoulos2012interference}.
 We find out precoding and receive filter matrices to minimize the rank of the interference matrix by solving the problem (6):
\begin{align}
	\underset{\boldsymbol{U}_{k},\boldsymbol{V}_{k},k=1,...,K}{\mathop{\min }}\,\sum\limits_{k=1}^{K}{rank(\boldsymbol{J}_{k})}\tag{6a}\\
	s.t.:rank(\boldsymbol{S}_{k})=d\tag{6b}
\end{align}  
 
\subsection{System Model for Relay-Aided MIMO Interference Channel}
 In this part, a system in which $X$ half-duplex AF relays aid $K$ transmitters and $K$ receivers is considered. The number of antennas at each receiver, transmitter and relay is denoted by $\mathop{M}_{r}$, $\mathop{M}_{t}$ and $\mathop{M}_{x}$, respectively. We assume that there is no direct link between transmitters and receivers. This system can be represented by ${{({{M}_{t}}*{{M}_{r}},d)}^{K}}+{ {M}_{x}}^{X}$. The channels between $k$-th transmitter and $x$-th relay, $x$-th relay and $k$-th receiver are denoted by $\mathop{\boldsymbol{H}}_{x,k}\in \mathop{\mathbb{C}}^{{M}_{x}*{M}_{t}}$ and $\mathop{\boldsymbol{G}}_{k,x}\in \mathop{\mathbb{C}}^{{M}_{r}*{M}_{x}}$, respectively.
 Since we use half-duplex AF relay, the transmission process contains two phases. In the first phase, the transmitters send their signal to the relays. Thus, $x$-th relay receives $boldsymbol{y}_{x,r}=\sum\limits_{k=1}^{K}\boldsymbol{H}_{x,k}\boldsymbol{V}_{k}\boldsymbol{s}_{k}+\boldsymbol{n}_{x}$. where $boldsymbol{V}_{k}$ is the precoding matrix at $k$-th transmitter and $\boldsymbol{n}_{x}$ represent the noise vector at $x$-th relay with zero mean and covariance $\sigma^{2}{\boldsymbol{I}_{_{M{}_{x}}}}$. In the second stage, the relays amplify their received signal and send it to the receivers. Therefore, after applying the linear receive filter ($\boldsymbol{U}_{k}$), $k$-th receiver observes: 
 \begin{align}
 	\boldsymbol{y}_{k}=\sum\limits_{x=1}^{X}\boldsymbol{U}_{k}^{H}\boldsymbol{G}_{k,x}\boldsymbol{W}_{x}\boldsymbol{R}_{x,k}+\boldsymbol{U}_{k}^{H}\sum\limits_{l=1,l\ne k}^{K}\sum\limits_{x=1}^{X}\boldsymbol{G}_{k,x}\boldsymbol{W}_{x}\boldsymbol{R}_{x,l}+\nonumber\\\boldsymbol\nonumber\boldsymbol{U}_{k}^{H}\sum\limits_{x=1}^{X}\boldsymbol{G}_{k,x}\boldsymbol{W}_{x}\boldsymbol{n}_{x}+\boldsymbol{U}_{k}^{H}\boldsymbol{f}_{k}\tag{7}
 \end{align}  
 where $\mathop{\boldsymbol{W}}_{x}\in \mathop{\mathbb{C}}^{{M}_{x}*{M}_{x}}$ and $\mathop{\boldsymbol{U}}_{k}\in \mathop{\mathbb{C}}^{{M}_{r}*{d}}$ are the processing matrix at $x-th$ relay and the linear receive matrix at $k-th$ receiver, respectively. Moreover, $\small\boldsymbol{R}_{x,l}=\boldsymbol{H}_{x,l}\boldsymbol{V}_{l}\boldsymbol{s}_{l}$ and $\boldsymbol{f}_{k}$ is the zero-mean white Gaussian noise with covariance $\sigma^{2}{\boldsymbol{I}_{_{M{}_{r}}}}$ at $k-th$ receiver. The second term in (7) is interference at each user produced by other transmitters. In addition, $n_{x}$ in the third term of (7) is enhanced by the relays. This issue cannot be overlooked because during the second stage of transmission, the power of this noise is scaled up with the power of the desire signal, and this provides serious obstacle in achieving higher DOF \cite{Troung2013}. Therefore, we should consider the interfrence and also enhanced noise in the design of approaches for AF relay-aided MIMO interference channels. The transmit power of $x-th$ relay and $k-th$ transmitter are ${PR}_{x}=\sum\limits_{k=1}^{K}tr(\boldsymbol{W}_{x}\boldsymbol{H}_{x,k}\boldsymbol{V}_{k}\boldsymbol{V}_{k}^{H}\boldsymbol{H}_{x,k}^{H}\boldsymbol{W}_{x}^{H})+\sigma^{2}{\boldsymbol{I}_{_{M{}_{x}}}}tr(\boldsymbol{W}_{x}\boldsymbol{W}_{x}^{H})$, ${PU}_{k}=tr(\boldsymbol{V}_{k}\boldsymbol{V}_{k}^{H})$, respectively.
 
 We define the signal and interference matrices $({\boldsymbol{S}_{k}},{\boldsymbol{J}_{k}})$ for all  $x=1,...,X$, $k= 1,...,K$ as follows:
 \begin{equation}
 {\boldsymbol{S}_{k}}=\sum\limits_{x=1}^{X}\boldsymbol{U}_{k}^{H}\boldsymbol{G}_{k,x}\boldsymbol{W}_{x}\boldsymbol{H}_{x,k}\boldsymbol{V}_{k}\tag{8a}
 \end{equation}
 \begin{equation}
 {\boldsymbol{J}_{k}}\triangleq \boldsymbol{U}_{k}^{H}[\{{\boldsymbol{G}_{k,x}}{\boldsymbol{W}_{x}}{\boldsymbol{H}_{x,l}}{\boldsymbol{V}_{l}}\}_{l=1,l\ne k,x=1}^{l=K,x=X}],{{\boldsymbol{J}}_{k}}\in {{\boldsymbol{C}}^{d\times[(K-1)d{X}]}}\tag{8b}
 \end{equation}  
 As we stated there is the enhanced noise in the system which cannot be ignored; therefore, we should minimize the power of this enhanced noise. Thus, optimization problem can be expressed as follows:
 \begin{align}
 \underset{\boldsymbol{U}_{k},\boldsymbol{V}_{k},\boldsymbol{W}_{x}, k=1,...,K, x=1,...,X}{\mathop{\min }}\,\sum\limits_{k=1}^{K}{rank(\boldsymbol{J}_{k})}+{N}\tag{9a}\\
 s.t.:{PR}_{x}={P}_{1}\tag{9b}\\
 {PU}_{k}={P}_{2}\tag{9c}\\
 rank(\boldsymbol{S}_{k})=d\tag{9d}
 \end{align}   
   where ${P}_{1}$, ${P}_{2}$ are amount of transmit power at each relay and transmitter, respectively. Moreover, ${N}$ is the sum power of the enhanced noise,  ${N}=\sum\limits_{k=1}^{K}\sum\limits_{x=1}^{X}\sigma^{2}tr{(\boldsymbol{U}_{k}^{H}}{\boldsymbol{G}_{k,x}}{\boldsymbol{W}_{x}{\small\boldsymbol{W}_{x}^{H}{\boldsymbol{G}_{k,x}^{H}}}}\boldsymbol{U}_{k})$. 
\section{PROPOSED METHODS}
In this section, two SDP-based rank minimization methods with ability to enhance the performance of IA in the presence of uncoordinated
interference are introduced. At the First, we introduce new objective function and study it effects on the performance of IA.
Then, we propose a new convex relaxation.

One of the most common convex optimization based heuristic approaches to solve rank minimization problem is nuclear norm/trace.
In the case of finding sparse vectors, this approach is transformed into $l_{1}$ norm minimization method. The advantage of such a method is
that can be solved efficiently \cite{fazel2002matrix}. Rank of a semidefinite symmetric matrix is equal to the number of its non-zero singular values and nuclear norm of this matrix is sum of the all singular values. In fact, by using the nuclear norm function in (6) it is wished that minimizing the sum of the singular values leads to decrease singular values and consequently, lower rank matrices are obtained. Both small and large positive singular values have equal effect on the rank of positive semidefinite matrices but, the $l_{1}$ norm approximation (nuclear norm) sets high emphasis on small singular values. In contrast, it puts the less weight on large singular values \cite{boyd2004convex}.

When perfect IA cannot be attainable and therefore rank of the interference matrix is not zero, difference between rank minimization
and nuclear norm minimization exposes. The amount of DOF is highly affected by weighting of singular values. This motivates us
to search for other relaxations to find out better singular values’ weighting approaches and achieve higher DOF.

In the following, we propose our optimization methods:
\subsection{${{l}_{2}}$ norm minimization with SDP constrained}
 Due to the fact that the rank function is intractable, the rank function should be approximated. In this section, we use ${{l}_{2}}$  norm approximation to solve RCRM problem (6). In comparison with ${{l}_{1}}$  norm approximation, large singular values get higher weights in ${{l}_{2}}$  norm approximation; consequently, ${{l}_{2}}$  norm approximation yields fewer large singular values than ${{l}_{1}}$  norm approximation \cite{boyd2004convex}. In the presence of uncoordinated interference, the singular values of interference matrix may increase. This trait encourages us to use ${{l}_{2}}$  norm approximation in the SDP-based rank minimization in order to obtain higher DOF. By using ${{l}_{2}}$  norm approximation, the problem (6) can be rewritten as follows:
\begin{equation} 
\underset{\boldsymbol{U}_{k},\boldsymbol{V}_{k},k=1,...,K}{\mathop{\min }}\,\sum\limits_{k=1}^{K}{tr(\boldsymbol{Y}_{k})}\tag{10a}\\
\end{equation}
\begin{align}
	s.t: &rank(\boldsymbol{S}_{k})=d \tag{10b}\\
	&\boldsymbol{Y}_{k}=\boldsymbol{J}_{k}^{H}\boldsymbol{J}_{k}\tag{10c}
\end{align} 
The problem (10) is not a convex optimization problem. To address this, (10c) should be relaxed. Hence, (10c) can be linearized by using the Schur complement, and can be expressed by the following linear matrix inequality (LMI) in (11c):
\begin{equation}
\min_{\boldsymbol{U}_k,\boldsymbol{V}_k, k=1,\ldots,K.}\sum_{k=1}^{K}tr(\boldsymbol{W}_k)\tag{11a}\\
\end{equation}
\begin{align}
	s.t: &rank(\boldsymbol{S}_k)=d\tag{11b}\\
	&\boldsymbol{W}_k\succeq \boldsymbol{0}_{[Kd+D_f]\times[Kd+D_f]}\tag{11c}\\
	&\boldsymbol{W}_k=\begin{pmatrix}
		\boldsymbol{I}_d & \boldsymbol{J}_k \\
		\boldsymbol{J}_k^H & \boldsymbol{Y}_k
	\end{pmatrix}\tag{11d}\\
	&\boldsymbol{S}_{k}\triangleq {\boldsymbol{U}}_{k}^{H}\boldsymbol{H}_{k,k}\boldsymbol{V}_{k}\tag{11e}\\
	&{{\boldsymbol{J}}_{k}}\triangleq\boldsymbol{U}_{k}^{H}\left[ \left\{ {{\boldsymbol{H}}_{k,l}}{{\boldsymbol{V}}_{l}} \right\}_{l=1,l\ne k}^{K}...\left\{ {{\boldsymbol{C}}_{k,f}}{{\boldsymbol{F}}_{f}} \right\}_{f=1}^{X} \right]\tag{11f}
\end{align}
Constraint (11b) is not a convex optimization problem constraint. To solve this problem the rank of signal matrices $({\boldsymbol{S}_{k}})$ should be constrained by a convex constraint. It can be easily shown that $\boldsymbol{S}_{k}$, which satisfies (12) is positive definite and also full rank:
\begin{equation} 
	\boldsymbol{S}_{k}-\gamma {\boldsymbol{I}}_{d}\succeq \boldsymbol{0}_{d*d}\tag{12}
\end{equation}
where $0<\gamma \ll 1$. Proof of (12): see Appendix (A)

 To expand the feasibility set of problem (11), we modify (12) by adding matrix $\boldsymbol{Z}$ as follows:
 \begin{equation}
 	\boldsymbol{S}_{k}+\boldsymbol{Z}-\gamma \times\boldsymbol{I}_{d}\succeq \boldsymbol{0}_{d*d}\tag{13}
 \end{equation}
 By using (13), new possible solutions are considered which have been disregarded in the previous RCRM-based IA algorithms. To analyze the effect of (13), we obtain dual form of the optimization problem (11) in the each iteration of alternating minimization approach as follows (note that we replace (11b) by (12) and we use alternating minimization approach \cite{Bert2011} to solve optimization problem (11). In each iteration of this method we fix one variable and solve optimization problem with respect to another variable; consequently, in the each iteration of alternating minimization approach, optimization problem (11) has one variable):  

\begin{align}
&\max (tr(\boldsymbol{A}_{1,k}))\tag{14a}\\
s.t:&{\boldsymbol{I}_{[Kd+{D_{f}}]}}-{\boldsymbol{A}_{2,k}}+{\boldsymbol{K}_{1,k}}{\boldsymbol{B}_{2,k}}\boldsymbol{K}_{2,k}^{T}&\nonumber\\+&{\boldsymbol{K}_{3,k}}{\small\boldsymbol{B}_{3,k}}\boldsymbol{K}_{3,k}^{T}={\boldsymbol{0}_{[Kd+{{D}_{f}}]\times[Kd+{{D}_{f}}]}}
\tag{14b}\\
&\boldsymbol{Q}\boldsymbol{B}_{1,k}+\boldsymbol{T}\boldsymbol{B}_{2,k}^{T}= \boldsymbol{0}_{M_r\times d}\tag{14c}\\
&\boldsymbol{B}_{1,k}-\boldsymbol{A}_{1,k}=\boldsymbol{0}_{d\times d}\tag{14d}\\
&\boldsymbol{A}_{1,k}\succeq \boldsymbol{0}_{d\times d}\tag{14e}\\
&\boldsymbol{A}_{2,k} \succeq \boldsymbol{0}_{[Kd+D_f]\times [Kd+D_f]}\tag{14f}
\end{align}
where $\boldsymbol{Q} ={\boldsymbol{H}_{k,k}}{\boldsymbol{V}_{k}}$, $\boldsymbol{T}=\left[\left\{ {\boldsymbol{H}_{k,l}}{\boldsymbol{V}_{l}} \right\}_{l=1,l\ne k}^{K}...\left\{ {{\boldsymbol{C}}_{k,f}}{{\boldsymbol{F}}_{f}} \right\}_{f=1}^{X} \right]$, and also ${\boldsymbol{K}_{1,k}}\in {{\mathbb{C}}^{[Kd+{D}_{f}]\times d}}$, ${\boldsymbol{K}_{2,k}}\in {{\mathbb{C}}^{[Kd+{D}_{f}]\times[(K-1)d+{D}_{f}]}}$ such that ${\boldsymbol{J}_{k}}=\boldsymbol{K}_{1,k}^{T}{\boldsymbol{W}_{k}}{\boldsymbol{K}_{2,k}}$. Lagrange multipliers associated with inequality constraints are ${\boldsymbol{A}_{1,k}}$ and ${\boldsymbol{A}_{2,k}}$. ${\boldsymbol{B}_{1,k}}$, ${\boldsymbol{B}_{2,k}}$, ${\boldsymbol{B}_{3,k}}$ are Lagrange multipliers associated with equality constraints.
Proof of (14): see Appendix (B).

\textbf{Proposition $1$}: Using the relaxation (13) instead of (12) does not change the constraints of the dual problem; however, the objective function of the dual problem of IA is changed to the following statement:
\begin{equation}
	\max \left( tr({\boldsymbol{A}_{1}}\times({\boldsymbol{I}_{d}}-\boldsymbol{Z})) \right)\tag{15}
\end{equation}
Proof: see Appendix (C).

 In the convex optimization, the optimal value of the dual problem is the lower bound on the optimal value of the primal problem. Thus,
 in general case there is a duality gap between optimal values of the primal and dual problems. On the other hand, when strong duality
 holds, duality gap is zero. This means that the optimal values of the primal and dual problems are equal. In the each iteration of alternating minimization approach our optimization problem is convex and SDP (note that we replace (11b) by (12)); therefore, strong duality holds \cite{boyd2004convex}. If $\boldsymbol{Z}$  satisfies that $tr({\boldsymbol{A}_{1}}\boldsymbol{Z})\ge 0$, it can be easily shown that the optimal value of the dual problem associated with the constraint (13) is smaller than the case in which $\boldsymbol{Z}=\boldsymbol{0}$. Therefore, adding matrix $\boldsymbol{Z}$ can cause more decrease in the rank of interference matrices than the problem of (11); consequently, matrix $\boldsymbol{Z}$ can enhance the performance of IA to achieve higher DOF. 

\textbf{Proposition $2$} (Proof: see Appendix D): Let the optimal values of problem (11) associated with the constraints (12) and (13) be ${{P}^{*}}$ and ${{P}^{*}}(\boldsymbol{Z})$, respectively. For all $\boldsymbol{Z}$, ${{P}^{*}}(\boldsymbol{Z})\ge {{P}^{*}}-tr(\boldsymbol{A}_{1}^{*}\boldsymbol{Z})$. In order to be as close as possible to the lower bound of ${{P}^{*}}(\boldsymbol{Z})$, $ \boldsymbol{Z}$ should be comparatively small. This means that, to obtain lower rank solution for the interference matrix we choose comparatively small values for the entries of matrix $\boldsymbol{Z}$ in a way that the condition $tr({\boldsymbol{A}_{1}}\boldsymbol{Z})\ge 0$ is satisfied.

As long as ${\boldsymbol{A}_{1}}$ is a positive semidefinite, $\boldsymbol{1}_{d}$ matrix satisfies the constraint $tr({\boldsymbol{A}_{1}}\boldsymbol{Z})\ge 0$ . Thus the matrix $\boldsymbol{Z}$ is chosen as follows:
\begin{equation}
\boldsymbol{Z}=z\times\boldsymbol{1}_{d}\tag{16}
\end{equation}

Although (13) does not guarantee that ${\boldsymbol{S}_{k}}$ is a positive definite matrix, our results show that by choosing sufficiently small $z$ our proposed relaxation method provides full rank ${\boldsymbol{S}_{k}}$. However, if ${\boldsymbol{S}_{k}}$ is not full rank matrix, we decrease $z$ step by step until a full rank matrix is attained. Furthermore, using $\boldsymbol{Z}$ matrix expands the feasibility set of the optimization problem, and this can enable RCRM-based approaches to obtain lower optimal values.
The whole procedure of alternating minimization approach is explained in Table~\ref{Table algorithm}.    

\begin{table}[t]
	\centering
	\caption{Algorithm for solving the optimization problem (11)}
	\begin{tabular}{lc}
		\hline
		1: choose arbitrary matrix ${\boldsymbol{U}_{k}}$, k=1,...,K\\
		2: for $n=1:n_{max}$ iteration\\ 
		3: fix ${\boldsymbol{U}_{k}}$ and solve optimization problem (11) $\to {\boldsymbol{V}_{k}}$\\
		4: fix ${\boldsymbol{V}_{k}}$ and solve optimization problem (11) $\to {\boldsymbol{U}_{k}}$\\
		5: if $rank({\boldsymbol{S}_{k}})\ne d$, $z\to z/2$ and jump to step 3\\
		6: end for \\
		\hline
	\end{tabular}
	\label{Table algorithm}
\end{table}
\subsection{Schatten-$p$ norm minimization}
In this section, we use Schatten-p norm as a surrogate of rank operator. We propose an iterative algorithm to solve optimization problem. The Schatten-p norm function for  $p\in (0,1]$ is defined as:
\begin{equation}
	{f}_{p}(\boldsymbol{A})=\sum\limits_{i>1}{({\delta }_{i}(\boldsymbol{A}))}^{p}\tag{17}
\end{equation}
Schatten-$p$-norm has special properties that make it good choice to approximate the rank operator. Firstly, for a positive semidefinite matrix, Schatten-$p$ norm equals to nuclear norm when $p=1$. Hence, nuclear norm is a special case of Schatten-$p$ norm. Secondly, if $p$ converges to zero, ${{f}_{p}}(\boldsymbol{A})\to rank(\boldsymbol{A})$. This indicates that Schatten-$p$ norm can find a low-rank solution when $p$ has small value \cite{ji2013beyond}. Therefore, we replace (6a) by (17). 

 To avoid the non-differentiability of (17), it can be changed to the following statement \cite{ge2011note}:
\begin{equation}
	\underset{\boldsymbol{U}_{k},\boldsymbol{V}_{k}}{\mathop{\min }}\,\sum\limits_{k=1}^{K}{\sum\limits_{i=1}^{d}{{({\delta }_{i}(\boldsymbol{J}_{k})+\xi )}^{p}}}\tag{18}\\
\end{equation}
where $\xi $ has small positive value.

 Now, we propose our iterative algorithm. Since (18) is not cost function of convex optimization problem, we should linearize it. In order to linearize (18) we use Taylor expansion. In the $l$-th iteration, the expansion of (18) is as follows:  
\begin{equation}
	{({{\delta }_{i}(\boldsymbol{J}_{k}^{l})+\xi )}^{p}}+(\frac{p}{{({\delta }_{i}(\boldsymbol{J}_{k}^{l})+\xi )}^{1-p}})({\delta }_{i}(\boldsymbol{J}_{k})-{\delta }_{i}(\boldsymbol{J}_{k}^{l}))\tag{19}\\
\end{equation}
Here, $\boldsymbol{J}_{k}^{l}\normalsize$ is fixed and can be removed. As a result, we have the objective function as:
\begin{equation}
	\underset{\boldsymbol{U}_{k},\boldsymbol{V}_{k}}{\mathop{\min }}\,\sum\limits_{k=1}^{K}{\sum\limits_{i=1}^{d}{\frac{p\times({{\delta }_{i}}({\boldsymbol{J}_{k}}))}{{{({{\delta }_{i}}(\boldsymbol{J}_{k}^{l})+\xi )}^{1-p}}}}}\tag{20}\\
\end{equation}
The objective function (20) can be written as (21a). Therefore, the optimization problem can be expressed as:

\begin{equation}
	\underset{\boldsymbol{U}_{k},\boldsymbol{V}_{k}}{\mathop{\min }}\,\sum\limits_{k=1}^{K}{{{\left\| \boldsymbol{G}_{k}^{l}{\boldsymbol{J}_{k}} \right\|}_{*}}}\tag{21a}\\
\end{equation}
\begin{equation}
	s.t:\boldsymbol{S}_{k}+\boldsymbol{Z}-\gamma \times\boldsymbol{I}_{d}\succeq\boldsymbol{0}_{d*d}\tag{21b}\\
\end{equation}
Proof of (21a): see Appendix (E).

where $\boldsymbol{G}_{k}^{l}={{\boldsymbol{\Lambda} }_{k}}\boldsymbol{\Phi} _{k}^{l}\boldsymbol{\Lambda} _{k}^{H}$ is the weight matrix and ${{\boldsymbol{\Lambda} }_{k}}$ is achieved by singular value decomposition of ${\boldsymbol{J}_{k}}$. $\boldsymbol{\Phi} _{k}^{l}\in {{\mathbb{C}}^{d\times d}}$ is a diagonal matrix that $ i$-th diagonal element is equal to $\frac{p}{{{({{\delta }_{i}}(\boldsymbol{J}_{k}^{l})+\xi )}^{1-p}}}$ .

We use the algorithm in Table~\ref{Table algorithmII} to solve the optimization problem (21). In this case in each iteration ($l$-th), the weight matrix $\boldsymbol{G}_{k}^{l+1}$ is updated according to the interference matrix $\boldsymbol{J}_{k}^{l}$.
\begin{table}[t]
	\centering
	\caption{Algorithm for solving the optimization problem (21)}
	\begin{tabular}{lc}
		\hline
		1: choose arbitrary matrix ${\boldsymbol{U}_{k}}$, k=1,...,K\\
		2: choose ${\boldsymbol{G}_{k}^{0}}={\boldsymbol{I}_{d}}$\\
		3: for $l=1:l_{max}$ iteration\\
		4: for $n=1:n_{max}$ iteration\\ 
		5: fix ${\boldsymbol{U}_{k}}$ and solve optimization problem (11) $\to {\boldsymbol{V}_{k}}$\\
		6: fix ${\boldsymbol{V}_{k}}$ and solve optimization problem (11) $\to {\boldsymbol{U}_{k}}$\\
		7: if $rank({\boldsymbol{S}_{k}})\ne d$, $z\to z/2$ and jump to step 3\\
		8: end for \\
		9: update ${\boldsymbol{G}_{k}^{l+1}}$\\
		10: end for\\
		\hline
	\end{tabular}
	\label{Table algorithmII}
\end{table}
\subsection{A Proposed Method for Relay-aided MIMO Interference Channel}
In this section, we propose a weighted-sum method to solve optimization problem (9). As a first step, we determine the precoding matrix ($\boldsymbol{V}_{k}$). However, before this we should apply some modifications. Firstly, in the optimization problem (9) we can use Schatten-p as a surrogate of rank function, and (9d) can be replaced by (13). Secondly, (9b) and (9c) do not represent the constraints of a convex optimization problem. We use Schur complement to turn them into the convex constraints. Therefore, we have optimization problem (22) (in the (22) we do not consider the second term of (9a) because $\boldsymbol{V}_{k}$ has not role in the enhanced noise). 
\begin{align}
\underset{\boldsymbol{V}_{k}, k=1,...,K}{\mathop{\min }}\,\sum\limits_{k=1}^{K}{tr(\boldsymbol{J}_{k})}\tag{22a}\\
s.t: \boldsymbol{Y}_{k}=\left( \begin{matrix}
\boldsymbol{I}_{{M}_{T}} & \boldsymbol{V}_{k}^{H}  \\
\boldsymbol{{V}_{k}} & \boldsymbol{{Y}_{kk}}  \\
\end{matrix} \right)\tag{22b}\\
\boldsymbol{Q}_{x,k}=\left( \begin{matrix}
\boldsymbol{I}_{d} & \boldsymbol{QW}_{x,k}^{H}  \\
\boldsymbol{{QW}_{x,k}} & \boldsymbol{{QA}_{x,k}}  \\
\end{matrix} \right)\tag{22c}\\
{\boldsymbol{Y}_{k}},{\boldsymbol{Q}_{x,k}}\succeq\boldsymbol{0}\tag{22d}\\
{tr}({\boldsymbol{Y}_{k}})={P}_{2}+{M}_{T}\tag{22e}\\
\sum\limits_{k=1}^{K}{tr({\boldsymbol{Q}_{x,k}})={{P}_{U}}}\tag{22f}\\
\boldsymbol{S}_{k}-\gamma {\boldsymbol{I}}_{d\times{d}}\succeq\boldsymbol{0}_{d\times{d}}\tag{22g}
\end{align}   
 where $\boldsymbol{Y}_{kk}=\boldsymbol{V}_{k}\boldsymbol{V}_{k}^{H}$ and $\boldsymbol{QW}_{x,k}=\boldsymbol{W}_{x}\small\boldsymbol{H}_{x,k}{{\boldsymbol{V}}_{k}}$, ${\boldsymbol{QA}_{x,k}}=\boldsymbol{QW}_{x,k}^{H}{\boldsymbol{QW}_{x,k}}$, ${{P}_{U}}={{P}_{1}}+d-tr({\boldsymbol{W}_{x}}\boldsymbol{W}_{x}^{H})$. The precoding matrix is determined by solving the convex optimization problem (22). After optimization to satisfy (9c), $\overset\frown{{\boldsymbol{V}_{k}}}$ is a new precoding matrix, which is equal to $\alpha {\boldsymbol{V}_{k}}$, where $\alpha =\sqrt{\frac{tr({\boldsymbol{Y}_{kk}})}{tr({\boldsymbol{V}_{k}}\boldsymbol{V}_{k}^{H})}}$. Now, we determine the processing  matrices at the relays and linear receive matrices. The second term of (9a) and constraint (9b) do not represent the objective function and the constraint of a convex optimization problem. To address these problems, we again use Schur Complement. Therefore, to obtain ($\boldsymbol{W}_{x}$) we have the optimization problem (23).
   \begin{align}
   \underset{\boldsymbol{W}_{x}, x=1,...,X}{\mathop{\min }}\,\sum\limits_{k=1}^{K}{tr(\boldsymbol{J}_{k})}+\sum\limits_{k=1}^{K}\sum\limits_{x=1}^{X}{\sigma^{2}tr(\boldsymbol{T}_{x,k})}\tag{23a}\\
   s.t: \boldsymbol{T}_{x,k}=\left( \begin{matrix}
   \boldsymbol{I}_{{M}_{T}} & \boldsymbol{TW}_{x,k}^{H}  \\
   \boldsymbol{{TW}_{x,k}} & \boldsymbol{{TA}_{x,k}}  \\
   \end{matrix} \right)\tag{23b}\\
   \boldsymbol{Q}_{x,k}=\left( \begin{matrix}
   \boldsymbol{I}_{d} & \boldsymbol{QW}_{x,k}^{H}  \\
   \boldsymbol{{QW}_{x,k}} & \boldsymbol{{QA}_{x,k}}  \\
   \end{matrix} \right)\tag{23c}\\
   \boldsymbol{C}_{x}=\left( \begin{matrix}
   \boldsymbol{M}_{x} & \boldsymbol{W}_{x}^{H}  \\
   \boldsymbol{{W}_{x}} & \boldsymbol{{CW}_{x}}  \\
   \end{matrix} \right)\tag{23d}\\
   \sum\limits_{k=1}^{K}{tr({\boldsymbol{Q}_{x,k}})+tr({\boldsymbol{C}_{x}})}={{P}_{1}}+d+{M}_{x}\tag{23e}\\
   {\boldsymbol{T}_{x,k}},{\boldsymbol{Q}_{x,k}},{\boldsymbol{C}_{x}}\succeq\boldsymbol{0}\tag{23f}\\
   \boldsymbol{S}_{k}-\gamma {\boldsymbol{I}}_{d\times{d}}\succeq \small\boldsymbol{0}_{d\times{d}}\tag{23g}
   \end{align}
   where ${\boldsymbol{TW}_{x,k}}=\boldsymbol{U}_{k}^{H}{\boldsymbol{G}_{k,x}}{\boldsymbol{W}_{x}}$, ${\boldsymbol{TA}_{x,k}}=\boldsymbol{TW}_{x,k}^{H}{\boldsymbol{TW}_{x,k}}$, ${\boldsymbol{CW}_{x}}=\boldsymbol{W}_{x}^{H}{\boldsymbol{W}_{x}}$, $\boldsymbol{QW}_{x,k}=\boldsymbol{W}_{x}\boldsymbol{H}_{x,k}{\overset\frown{\boldsymbol{V}}_{k}}$. To obtain the linear receive matrix  ($\boldsymbol{U}_{k}$), we have (24).
   \begin{align}
   \underset{\boldsymbol{U}_{k}, k=1,...,K}{\mathop{\min }}\,\sum\limits_{k=1}^{K}{tr(\boldsymbol{J}_{k})}+\sum\limits_{k=1}^{K}\sum\limits_{x=1}^{X}{\sigma^{2}tr(\boldsymbol{T}_{x,k})}\tag{24a}\\
   s.t: \boldsymbol{T}_{x,k}=\left( \begin{matrix}
   \boldsymbol{I}_{{M}_{T}} & \boldsymbol{TW}_{x,k}^{H}  \\
   \boldsymbol{{TW}_{x,k}} & \boldsymbol{{TA}_{x,k}}  \\
   \end{matrix} \right)\tag{24b}\\
   {\boldsymbol{T}_{x,k}}\succeq\boldsymbol{0}\tag{24c}\\
   \boldsymbol{S}_{k}-\gamma {\boldsymbol{I}}_{d\times{d}}\succeq\boldsymbol{0}_{d\times{d}}\tag{24d}
   \end{align}
   It is noted that, another SDP-relaxation which is named as Rank-one relaxation has been used in \cite{Troung2013}. The solution of the optimization problem relaxed by Rank-one relaxation must be rank-one. But there is not any guaranty that solving the optimization problem (associated with rank-one relaxation) by available software package provides rank-one solution. Therefore, usually in such a problem, to obtain a rank-one solution, the rank reduction procedures are employed. These rank reduction procedures have computational burden; however, the relaxation used (Shur Complement) in this paper does not need rank reduction procedures.
   
   Due to the fact that the optimization problems (23), (24) are bi-objective, we should find pareto (efficient) point. The multi-objective optimization problem can be stated as follows. 
    \begin{align}
    	\underset{x}{\mathop{\min }}\,{{f}_{1}}(x)+{{f}_{2}}(x)+...+{{f}_{m}}(x)\tag{25}
    \end{align}
    Since there is not usually a single point (a global optimum) that can optimize every function, in the multi-objective optimization we should find pareto (efficient) points. A feasible point is strong pareto if there is not any other better feasible point. In addition, a feasible point is local (weak) pareto if in its neighborhood there is not better feasible point. However, existing methods can just guarantee that the local pareto points are found by using them \cite{Das1998}. Indeed, there is a trade-off between local pareto points which denotes that in order to do better on the some objectives how much worse we can do on another objective \cite{boyd2004convex}.
    Weighted-sum method is a technique for finding pareto optimal points. In this method, a constant weight is allotted to each objective; for instance, if we want one objective to be small, we should take a big weight. However, the weighted-sum method cannot find the whole pareto solutions, and its solutions are not uniformly distributed \cite{Kim2005}. In this paper, we propose a modified weighted-sum method to find more pareto optimal points inspired from previous work \cite{Kim2005}. Therefore, we can reformulate (23a), (24a) as (26) (it is noted that in this method we determine $\boldsymbol{U}_{k}$, $\boldsymbol{W}_{x}$ after obtaining $\boldsymbol{V}_{k}$ by solving problem (22)).
    \begin{align}
    	{\mathop{\min }}\,\sum\limits_{k=1}^{K}{tr({w}_{1}\boldsymbol{J}_{k})}+\sum\limits_{k=1}^{K}\sum\limits_{x=1}^{X}\sigma^{2}\small{tr({w}_{2}\boldsymbol{T}_{x,k})}\tag{26}
    \end{align}
    where ${w}_{1}, {w}_{2}$ are weights and $\sum\limits_{i=1}^{2}{{{w}_{i}}}=1$, ${w}_{i}>{0}$. The procedure of modified weighted-sum method is stated in Table~\ref{Table algorithmIII}.
    \begin{table}[t]
    	\centering
    	\caption{Modified Weighted-sum Method}
    	\begin{tabular}{lc}
    		\hline
    		1: choose ${w}_{1}=\alpha_{1}$, ${w}_{2}=1-\alpha_{1}$ and solve optimization problems (23), (24)\\
    		2: choose ${w}_{2}=\alpha_{1}$, ${w}_{1}=1-\alpha_{1}$ and solve optimization problems (23), (24)\\
    		3: choose ${w}_{1}=\alpha_{2}$, ${w}_{2}=1-\alpha_{2}$ ($\alpha_{2}<\alpha_{1}$)\\
    		4: To find more optimal points the constraints (27a), (27b) are added\\\ to optimization problems (23), (24)\\ 
    		5: solve optimization problems (23), (24)\\
    		6: choose ${w}_{2}=\alpha_{2}$, ${w}_{1}=1-\alpha_{2}$ ($\alpha_{2}<\alpha_{1}$)\\
    		7: solve optimization problems (23), (24)\\
    		\hline
    	\end{tabular}
    	\label{Table algorithmIII}
    \end{table}
	   In the proposed method, to find more optimal points we impose extra constraints on the problems (23), (24) as follows:
	   \begin{align}
	   tr({{w}_{1}}{\boldsymbol{J}_{k}})\le {{Z}_{1}}^{x}-\delta 1\tag{27a}\\
	   tr({{w}_{2}}{\boldsymbol{T}_{x,k}})\le {{Z}_{2}}^{y}-\delta 2\tag{27b}
	   \end{align}
	   It is useful to note that X and Y axes are illustrated amounts of ${tr({w}_{1}\boldsymbol{J}_{k})}$ and ${tr({w}_{2}\boldsymbol{T}_{x,k})}$, respectively. Moreover, $\delta 1=\delta{\cos (\theta )}$, $\delta 2=\delta{\sin (\theta )}$ and $\theta ={{\tan }^{-1}}(-\frac{Z_{1}^{y}-Z_{2}^{y}}{Z_{1}^{x}-Z_{2}^{x}})$. $\delta$ is a parameter chosen arbitrarily and ${Z}_{1}^{x}$ and ${Z}_{1}^{y}$ are the position of $tr({l{w}_{1}}{\boldsymbol{J}_{k}})$ and ${tr({w}_{2}\boldsymbol{T}_{x,k})}$ (calculated by  ${w}_{1}=\alpha_ {1}, {w}_{2}=1-\alpha_ {1}$), respectively. Furthermore, ${Z}_{2}^{x}$ and ${Z}_{2}^{y}$ are the position of $tr({{w}_{1}}{\boldsymbol{J}_{k}})$ and $tr({{w}_{2}}{\boldsymbol{T}_{x,k}})$ (calculated by ${w}_{1}=1-\alpha_{1}, {w}_{2}=\alpha_ {1}$), respectively. In the Table~\ref{Table algorithmIII}, we stop algorithm arbitrarily and similar to previous sections we use alternating minimization to solve optimization problems (23), (24). At the end of procedure, to satisfy (6b) the new relay processing matrix ($\overset\frown{{\boldsymbol{W}_{x}}}$) is $\overset\frown{{\boldsymbol{W}_{x}}}=\beta {\boldsymbol{W}_{x}}$, where $\beta =\sqrt{\frac{{{P}_{1}}}{tr({{{\boldsymbol{W}}}_{x}}({\boldsymbol{H}_{x,k}}\overset\frown{{\boldsymbol{V}_{k}}}\overset\frown{\boldsymbol{V}_{k}^{H}}\boldsymbol{H}_{x,k}^{H}+{{\sigma }^{2}}{\boldsymbol{I}_{{{M}_{r}}}}){{\boldsymbol{W}_{x}^{H}}})}}$, and we choose relay processing and linear receive matrices which yield the highest sum rate. Table~\ref{Table algorithm4} represents the whole procedure of our proposed method to find $\boldsymbol{V}_{K}$, $\boldsymbol{W}_{x}$ and $\boldsymbol{U}_{K}$.   
	   \begin{table}[t]
	   	\centering
	   	\caption{Algorithm for Finding $\boldsymbol{V}_{K}$, $\boldsymbol{W}_{x}$ and $\boldsymbol{U}_{K}$}
	   	\begin{tabular}{lc}
	   		\hline
	   		1: choose arbitrary matrices $\boldsymbol{W}_{x}$ and $\boldsymbol{U}_{K}$\\
	   		2: for $n=1:n_{max}$ iteration\\
	   		3: fix $\boldsymbol{W}_{x}$ and $\boldsymbol{U}_{K}$ and determine $\boldsymbol{V}_{K}$ by solving (23)\\
	   		4: fix $\boldsymbol{V}_{K}$ determine $\boldsymbol{W}_{x}$ and $\boldsymbol{U}_{K}$ by using &\\ the algorithm in the table Table~\ref{Table algorithmIII} \\
	   		5: end for\\
	   		\hline
	   	\end{tabular}
	   	\label{Table algorithm4}
	   \end{table}
\section{SIMULATION RESULTS}
In order to evaluate the performance of the proposed methods, numerical results are reported. We run simulations by using MATLAB toolbox CVX \cite{grant2008cvx}. Noise power level is considered as ${{\sigma }^{2}}=1$. The algorithms are performed over 200 channel realizations, where channel elements are drawn i.i.d from a Gaussian distribution with mean zero and variance 1.  Matrix $\boldsymbol{Z}$ is set to ${\boldsymbol{{1}}_{d}}$. Our methods for MIMO interference channel are compared with nuclear norm minimization method \cite{papailiopoulos2012interference}, the SINR maximization \cite{peters2011cooperative}, Log-det heuristic \cite{du2013reweighted} and the interference leakage minimization \cite{peters2009interference} approaches.

 In addition, the modified weighted-sum method is compared with the Leakage minimization approach \cite{Troung2013} and the weighted sum mean square (WMSE) minimization method \cite{Troung2013}. We choose $\delta$=0.01. In the modified weighted-sum method, at the first step we choose ${w}_{1}=0.8, {w}_{2}=0.2$ and then we reverse weights. At the next step, we choose ${w}_{1}=0.7, {w}_{2}=0.3$ then we reverse weights (at this step and next step, we solve optimization problems (23), (24) which have constraint (27)). Finally, we set ${w}_{1}=0.55, {w}_{2}=0.45$, and we choose the best matrices ($\boldsymbol{W}_{x}, \boldsymbol{U}_{k}$) which yield the highest sum rate. Besides, similar to \cite{Troung2013} we choose $PR_{x}=PU_{k}$. Due to the fact that weighting impacts on the achieved sum-rate, we try to use different weighting to find the best matrices.   

In Fig. 1 and Fig. 2, we consider ${{(10\times 6,4)}^{3}}$ MIMO interference system with one uncoordinated source (with fixed power 0dB).
 Fig. 1 depicts average multiplexing gain versus rank of uncoordinated interference. The rank of uncoordinated interference varies from $1$ to $4$. As it can be seen in the Fig. 1, when the rank of uncoordinated interference reaches to $3$, other methods cannot provide any average multiplexing gain while our proposed method has considerably better performance.  
\begin{figure}[t]
\begin{center}
\scalebox{0.8}
{\includegraphics{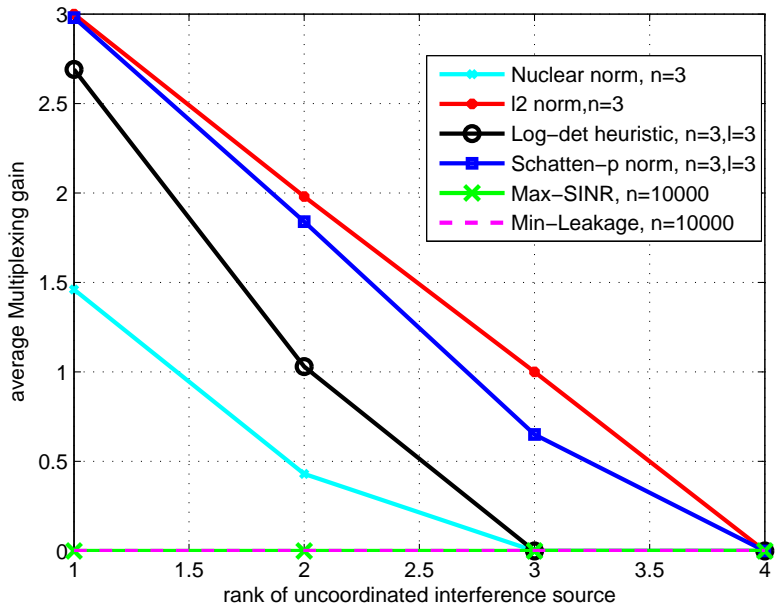}}
\end{center}
\caption{ Average multiplexing gain versus the rank of the uncoordinated interference source (${{P}}=30dB$)}
\label{fig:1}
\end{figure}

 Fig. 2 presents the average sum rate versus $P$ when the rank of uncoordinated interference is one. From Fig. 2, it is clearly seen that {${{l}_{2}}$ norm minimization method outperforms the other ones when SNR is higher than 20dB. Furthermore, the proposed methods can achieve higher sum rate than other approaches when $P$ is increased. Expanding the feasibility set of the optimization problem is one of the reasons for such results especially when uncoordinated interferences have strong effect and can considerably degrade the performance of IA. 
 	
 	In Fig. 3, we use ${{(10\times 6,4)}^{3}}$ MIMO interference system with two sources of uncoordinated interference. The rank of the first
 	uncoordinated source varies from 1 to 3, and the rank of the second source is fixed to 1. Fig. 3 displays average multiplexing gain versus
 	rank of uncoordinated interference sources (note that, in Fig. 3, the rank of the first source is just placed on the x-axis). In Fig. 3, it can be observed that the proposed methods noticeably outperform the other approaches. For example, when the rank of the first uncoordinated source is 2 and the second one is 1, other methods cannot obtain any average multiplexing gain, but our proposed method improves the performance of IA significantly.
 	
 	Fig. 4 represents the average achievable sum rate of ${{(6\times 4,2)}^{4}}$ MIMO interference system with one source of uncoordinated interference. As it can be seen the proposed methods outperform the other methods especially at higher SNR. 
 	
 	Table~\ref{Table Comparsion} represents average computation time of the proposed methods and the other algorithms. In this table, NN denotes Nuclear norm minimization method. The first and the second row of the table ~\ref{Table Comparsion} belongs to ${{(10\times 6,4)}^{3}}$ and ${{(6\times 4,2)}^{4}}$ MIMO interference system, respectively. Compared to other algorithms, $l_2$ norm minimization outperforms the other methods in terms of computation time. Indeed, in some cases such as ${{(6\times 4,2)}^{4}}$ MIMO interference system at low SNR our proposed methods and some approaches such as Max-SINR achieve comparatively equal sum-rate; however, our proposed methods especially $l_{2}$ norm minimization take less computation time than the other approaches.    
 	\begin{table}[t]
 		\centering
 		\caption{Average Computation Time}
 		\begin{tabular}{|l|c|c|c|c|c|c|}
 			\hline
 			method &$l_2$ norm  &NN  &Leak  &Log  &SINR&Schatten  \\\hline
 			Time(S) &3.95  &8.47  &12.15  &25.45  &85.89&23.97    \\\hline   
 			Time(S) &3.77  &5.8  &16.94  &16.02  &45.5&14.74    \\\hline  
 		\end{tabular}
 		\label{Table Comparsion}
 	\end{table}
 	
 		
 	Fig. 5 presents the average achievable sum rate versus $\frac{{{P}_{1}}}{{{\sigma }^{2}}}$. We depicts Fig. 5 for a ${{(2\times 2,2)}^{3}}+{{3}^{2}}$ system. As it can be seen from Fig. 5 the proposed method can outperform the other methods. Fig. 6 shows the average achievable sum rate versus the $\frac{{{P}_{1}}}{{{\sigma }^{2}}}$ for a ${{(4\times 2,2)}^{4}}+{{2}^{4}}$ system. It is clearly seen that the proposed methods can improve the average sum rate remarkably. The reason for this improvements is that in the rank minimization approach we try to minimize the dimensions occupied by interference signal which leads to obtain higher DOF. Increasing DOF can cause achieving higher sum rate particularly at medium and also relatively high $\frac{{{P}_{1}}}{{{\sigma }^{2}}}$. In addition, the Leakage minimization approach minimizes the energy of leakage interference. As a result, it generates low energy solutions; however, the solutions of the rank minimization approach are low rank which can yield higher DOF. Moreover, at medium and high $\frac{{{P}_{1}}}{{{\sigma }^{2}}}$, some transmitters have much smaller rates than the others in the WMSE approach. This cause a crucial obstacle to obtain higher sum rate at those $\frac{{{P}_{1}}}{{{\sigma }^{2}}}$. 
 	            
\begin{figure}[t]
\begin{center}
\scalebox{0.8}
{\includegraphics{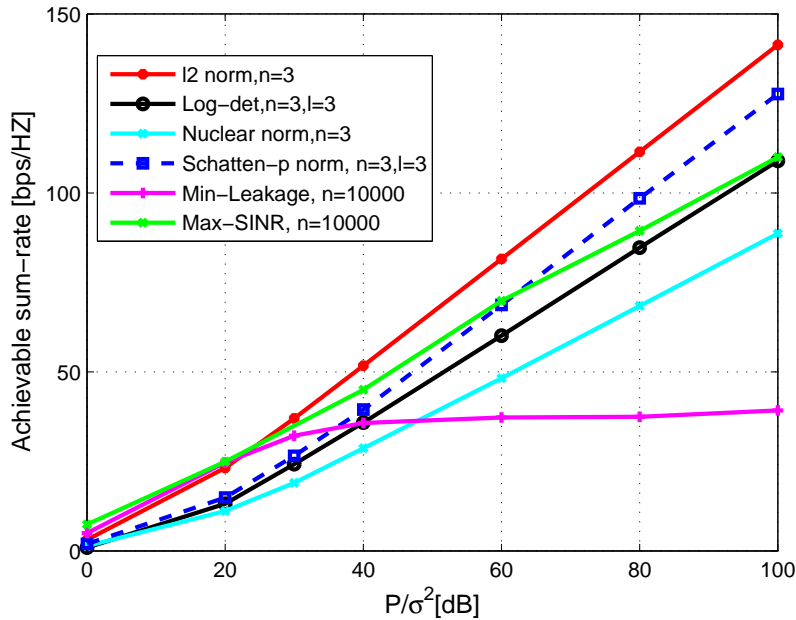}}
\end{center}
\caption{ Average sum rate for ${{(10\times 6,4)}^{3}}$ MIMO interference channel}
\label{fig:2}
\end{figure}

\begin{figure}[t]
	\begin{center}
		\scalebox{0.8}
		{\includegraphics{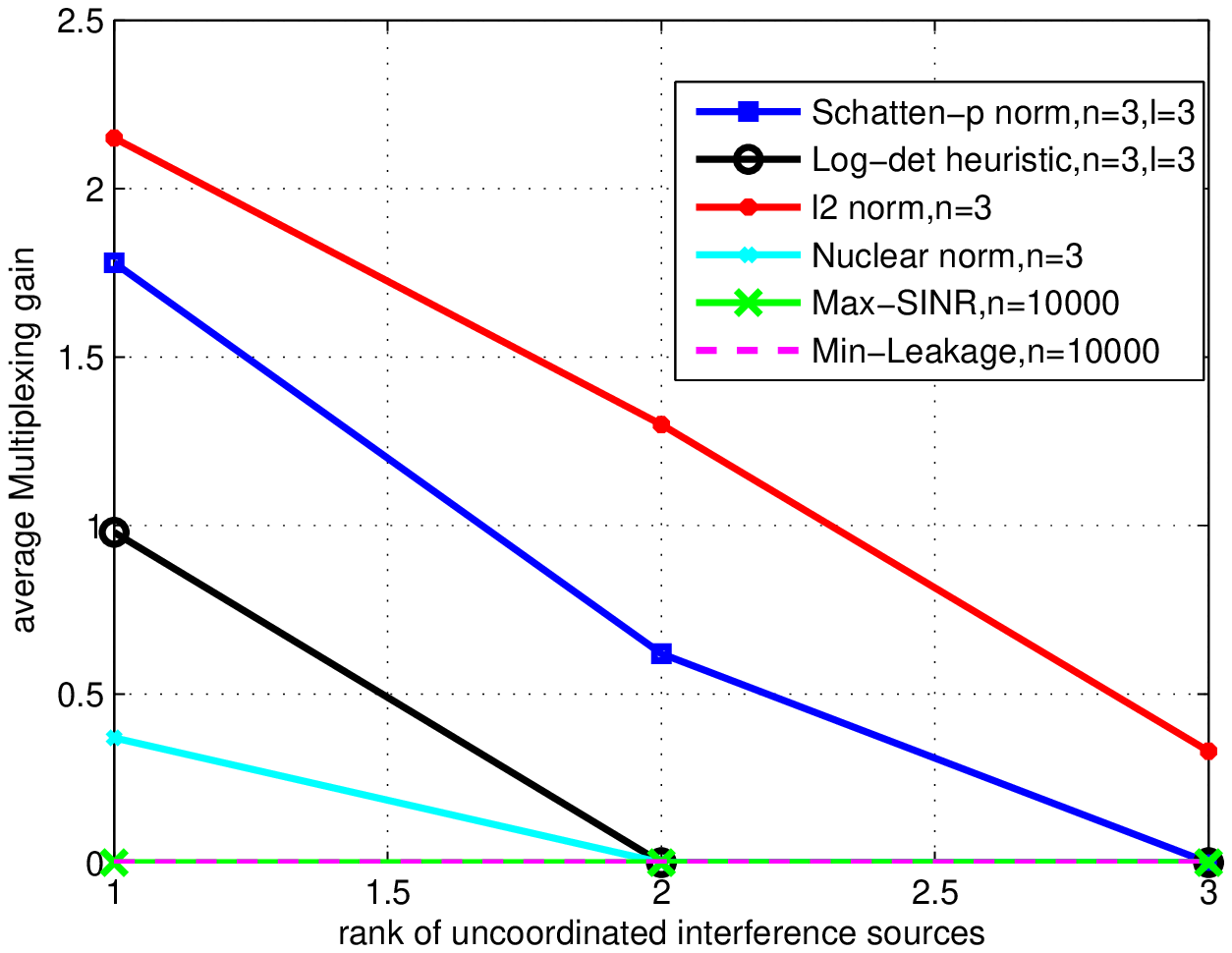}}
	\end{center}
	\caption{Average multiplexing gain versus the rank of the uncoordinated interference sources (${{P}}=30dB$)}
	\label{fig:3}
\end{figure}

\begin{figure}[t]
	\begin{center}
		\scalebox{0.8}
		{\includegraphics{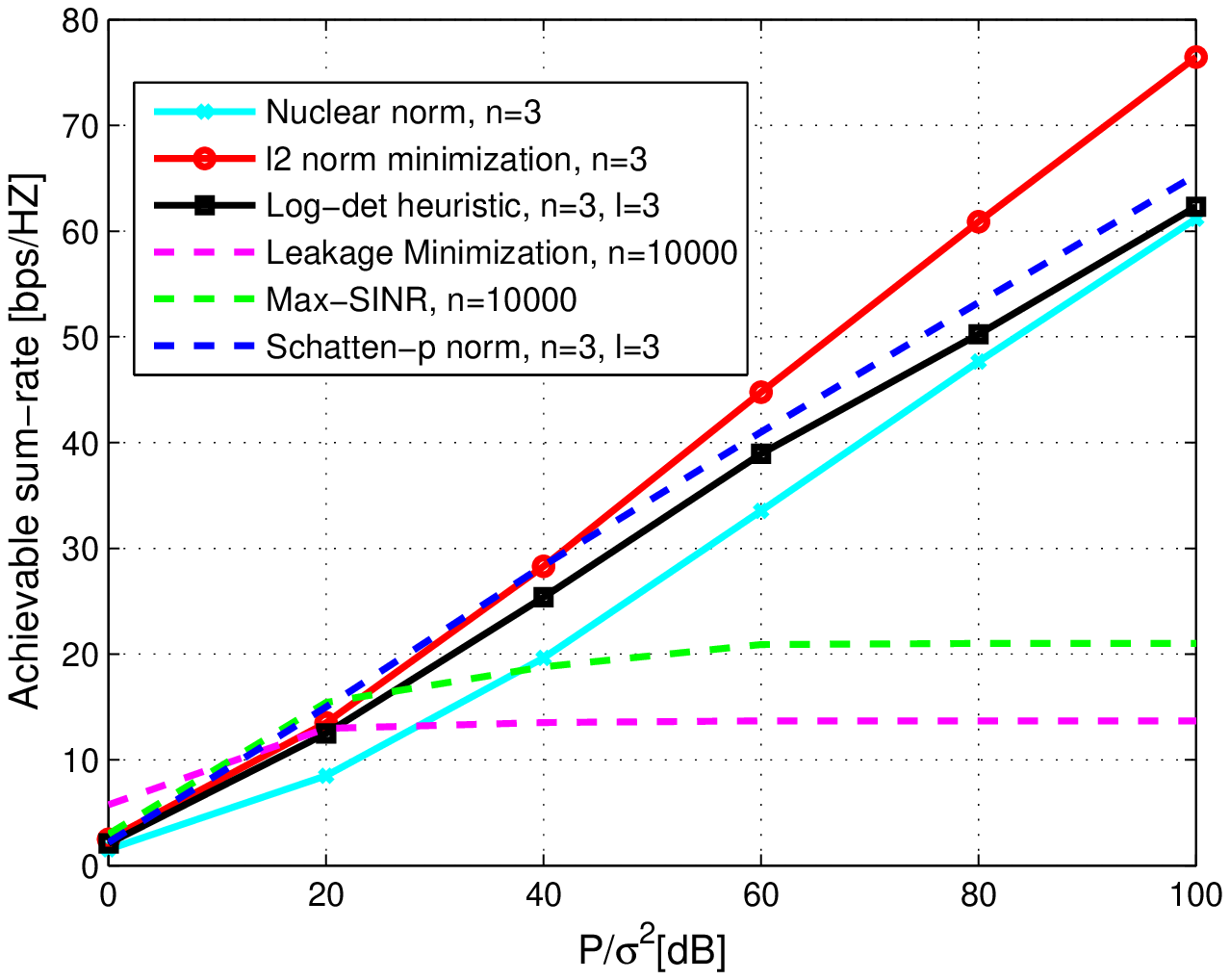}}
	\end{center}
	\caption{ Average sum rate for ${{(6\times 4,2)}^{4}}$ MIMO interference channel}
	\label{fig:4}
\end{figure}
\begin{figure}[t]
	\begin{center}
		\scalebox{0.8}
		{\includegraphics{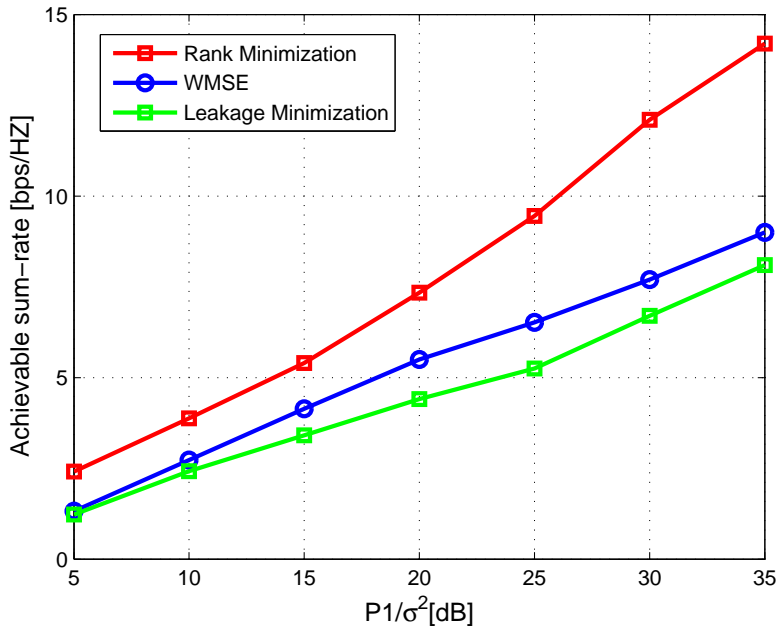}}
	\end{center}
	\caption{ Average sum rate for ${{(2\times 2,2)}^{3}}+{{3}^{2}}$ MIMO interference channel}
	\label{fig:5}
\end{figure} 
\begin{figure}[t]
	\begin{center}
		\scalebox{0.8}
		{\includegraphics{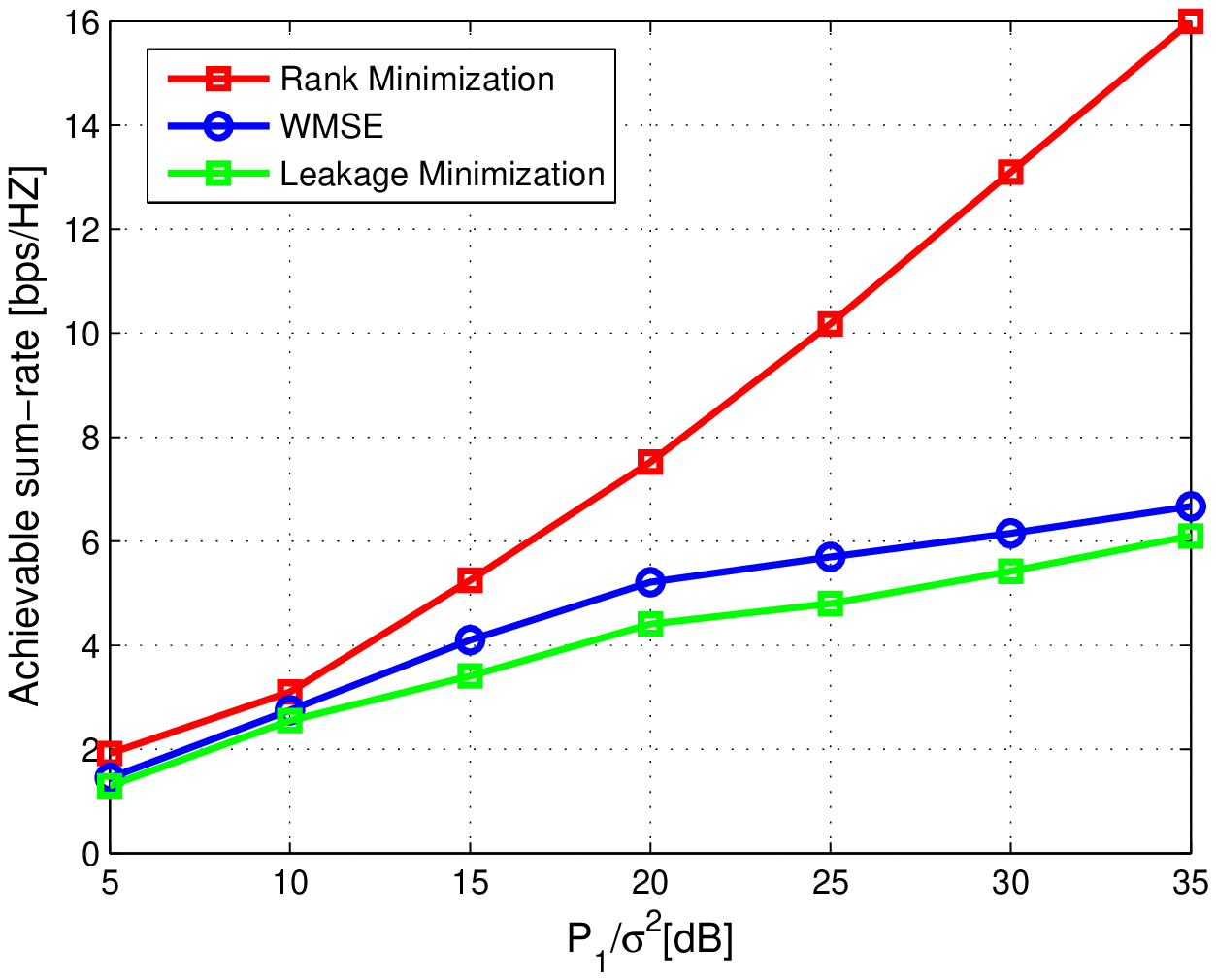}}
	\end{center}
	\caption{ Average sum rate for ${{(4\times 2,2)}^{4}}+{{2}^{4}}$ MIMO interference channel}
	\label{fig:6}
\end{figure} 
\section{CONCLUSION}
In this paper, we propose two rank minimization methods to improve the performance of IA in the MIMO interference channel
with uncoordinated interference sources. In addition, we expand the feasibility set of rank minimization problem by proposing a new convex relaxation, which can reduce the optimal value of our optimization problem, in order to achieve
higher DOF when the rank of interference matrix is large. Besides, we generalize our proposed methods to relay-aided MIMO interference channel ,and a modified weighted-sum method is proposed which can find more optimal points. Simulation results show that our proposed methods can achieve noticeably higher number of interference-free dimensions and sum rate compared to recently proposed approaches in IA framework.

\section{Appendix A}
A symmetric Matrix $\boldsymbol{M}$ is positive definite if the scalar $\boldsymbol{a}^{T}\boldsymbol{M}\boldsymbol{a}$ is positive for every vector $\boldsymbol{a}$ ($\boldsymbol{a}$ is non-zero column vector). Since (9) is positive semidefinite, we have: 
\begin{equation}
\boldsymbol{a}^{T} (\boldsymbol{S}_{k}-\boldsymbol{I}_{d})\boldsymbol{a}\ge 0\tag{28}
\end{equation}
According to (28):
\begin{equation}
\boldsymbol{a}^{T}\boldsymbol{S}_{k}\boldsymbol{a} \ge \boldsymbol{a}^{T} (\boldsymbol{I}_{d})\boldsymbol{a} > 0\tag{29}
\end{equation}
According to (29), and Due to the fact that $\boldsymbol{I}_{d}$ is positive definite matrix, $\boldsymbol{S}_{k}$ is also positive definite and full rank.
\section{Appendix B}
This section presents the derivation of dual problem (11) (in the each iteration of alternating minimization approach) associated with linear receive matrix. In the each iteration of alternating minimization approach we have optimization problem with one variable. To obtain the dual of problem (11), we use Lagrange dual function. The Lagrangian associated with problem (11) is expressed as follows: 
\begin{align}
L(\boldsymbol{U},\boldsymbol{A},\boldsymbol{B})&=\sum_{k=1}^K tr(\boldsymbol{W}_k)+\sum_{k=1}^K tr(\boldsymbol{A}_{1,k}(\boldsymbol{-S}_k+\boldsymbol{I}_d))+\nonumber\\&\sum_{k=1}^K tr(\boldsymbol{B}_{1,k}(\boldsymbol{S}_k-\boldsymbol{U}_k^H\boldsymbol{Q})+\sum_{k=1}^K tr(\boldsymbol{A}_{2,k}(\boldsymbol{-W}_k))\nonumber\\&+\sum_{k=1}^K tr(\boldsymbol{B}_{3,k}\boldsymbol{K}_{3,k}^{T}\boldsymbol{W}_{k}\boldsymbol{K}_{3,k})
\nonumber\\&+\sum_{k=1}^K tr(\boldsymbol{B}_{2,k}(\boldsymbol{J}_k-\boldsymbol{U}_k^H{\boldsymbol{T}}))\tag{30}
\end{align}
where $\boldsymbol{T}=\left[\left\{ {\boldsymbol{H}_{k,l}}{\boldsymbol{V}_{l}} \right\}_{l=1,l\ne k}^{K}...\left\{ {{\boldsymbol{C}}_{k,f}}{{\boldsymbol{F}}_{f}} \right\}_{f=1}^{X} \right]$, $\boldsymbol{Q} ={\boldsymbol{H}_{k,k}}{\boldsymbol{V}_{k}}$. By using (11d), we can represent interference matrix ($\boldsymbol{J}_{k}$) based on ($\boldsymbol{W}_{k}$) as follows ${\boldsymbol{J}_{k}}=\boldsymbol{K}_{1,k}^{T}{\boldsymbol{W}_{k}}{\boldsymbol{K}_{2,k}}$ where ${\boldsymbol{K}_{1,k}}\in {{\mathbb{C}}^{[Kd+{D}_{f}]\times d}}$, ${\boldsymbol{K}_{2,k}}\in {{\mathbb{C}}^{[Kd+{D}_{f}]\times[(K-1)d+{D}_{f}]}}$.  Therefore, we replace ($\boldsymbol{J}_{k}$) by ($\boldsymbol{W}_{k}$) in (30). In addition, with respect to (11d) $\boldsymbol{W}_{k}(1:d,1:d)=\boldsymbol{I}_{d}$, and we should consider this constraint to derivation of the dual problem. In fact, the matrices $\boldsymbol{K}_{3,k}$ satisfy the following statement.
\begin{align}
\boldsymbol{K}_{3,k}^{T}{\boldsymbol{W}_{k}}{\boldsymbol{K}_{3,k}}=\boldsymbol{I}_{d}\tag{31}
\end{align}
As long as (30) is linear function in order to prevent that it becomes unbounded below the following constraints should be hold.  Consequently, the cost function of dual problem is given by:
\begin{align}
&\max (tr(\boldsymbol{A}_{1,k}))\tag{32a}\\
s.t:&{\boldsymbol{I}_{[Kd+{D_{f}}]}}-{\boldsymbol{A}_{2,k}}+{\boldsymbol{K}_{1,k}}{\boldsymbol{B}_{2,k}}\boldsymbol{K}_{2,k}^{T}&\nonumber\\+&{\boldsymbol{K}_{3,k}}{\boldsymbol{B}_{3,k}}\boldsymbol{K}_{3,k}^{T}={\boldsymbol{0}_{[Kd+{{D}_{f}}]\times[Kd+{{D}_{f}}]}}
\tag{32b}\\
&\boldsymbol{Q}\boldsymbol{B}_{1,k}+\boldsymbol{T}\boldsymbol{B}_{2,k}^{T}=\boldsymbol{0}_{M_r\times d}\tag{32c}\\
&\boldsymbol{B}_{1,k}-\boldsymbol{A}_{1,k}=\boldsymbol{0}_{d\times d}\tag{32d}\\
&\boldsymbol{A}_{1,k}\succeq \boldsymbol{0}_{d\times d}\tag{32e}\\
&\boldsymbol{A}_{2,k} \succeq \boldsymbol{0}_{[Kd+D_f]\times [Kd+D_f]}\tag{32f}
\end{align}
Lagrange multipliers associated with inequality constraints are ${\boldsymbol{A}_{1,k}}$ and ${\boldsymbol{A}_{2,k}}$. ${\boldsymbol{B}_{1,k}}$, ${\boldsymbol{B}_{2,k}}$, ${\boldsymbol{B}_{3,k}}$ are Lagrange multipliers associated with equality constraints.
\section{Appendix C}
Adding matrix $\boldsymbol{Z}$ to constraint (13) changes the second term of (30) as follows $\sum_{k=1}^{K}tr(\boldsymbol{A}_{1,k}(\boldsymbol{-S}_{k}+\boldsymbol{I}_{d}-\boldsymbol{Z}))$. Due to the fact that matrix $\boldsymbol{Z}$ does not depend on any variables of optimization problem, it has not any role in the constraints (32); thus, the cost function of dual problem is changed as follows:
\begin{align}
\max \left( tr({\boldsymbol{A}_{1,k}}\times({\boldsymbol{I}_{d}}-\boldsymbol{Z})) \right)\tag{33}
\end{align} 
\section{Appendix D}
In this section, we want to propose a criterion to determine matrix $\boldsymbol{Z}$. Suppose that the optimal value of the problem (11) in the case that $\boldsymbol{Z}=0$ is ${P}^*$. Let ${P}^* (\boldsymbol{Z})$ denotes the optimal value of the problem (11) when we have a non-zero Z. As it is mentioned before, based on strong duality, the optimal value of both dual and original problem is equal. Now, let $\boldsymbol{A}_{1}^{0}$ be the optimal matrix of the problem (11) in the case that $\boldsymbol{Z}=0$. Similarly, suppose that $\boldsymbol{A}_{1}^{\boldsymbol{Z}}$ be the optimal matrix of the problem (11) when we have a non-zero $\boldsymbol{Z}$. When $\boldsymbol{Z}$=0, the maximum value of $tr(\boldsymbol{A}_{1} )$ associated to the constraints (11b) to (11f) is $tr(\boldsymbol{A}_{1}^0 )$. When we have a non-zero $\boldsymbol{Z}$, in comparison with the case that $\boldsymbol{Z}=0$ the constraints of the problem (11) remain unchanged but the objective function changes as $tr(\boldsymbol{A}_{1} (\boldsymbol{I}-\boldsymbol{Z}))$. Therefore, we can conclude that:
\begin{align}
&tr(\boldsymbol{A}_{1}^{0})\ge{tr(\boldsymbol{A}_{1}^{\boldsymbol{Z}})}\tag{34}
\end{align}
In fact, when we have a non-zero $\boldsymbol{Z}$, the objective is not to maximize the $tr(\boldsymbol{A}_{1})$ according to the constraints (11b) to (11f). we have:
\begin{align}
&{P}^{*}=tr(\boldsymbol{A}_{1}^0 )\tag{35a}\\
&{P}^{*} (Z)=tr(\boldsymbol{A}_{1})^{\boldsymbol{Z}}-tr(\boldsymbol{A}_{1})^{\boldsymbol{Z}} \boldsymbol{Z}\tag{35b}
\end{align}
Then we can write the following equation:
\begin{align}
&{P}^{*}-{P}^{*} (\boldsymbol{Z})=tr(\boldsymbol{A}_{1}^{0} )-tr(\boldsymbol{A}_{1})^{\boldsymbol{Z}}+tr(\boldsymbol{A}_{1}^{\boldsymbol{Z}}\boldsymbol{Z})\tag{36}
\end{align}
As long as $tr(\boldsymbol{A}_{1}^{0})-\boldsymbol{A}_{1}^{\boldsymbol{Z}}\ge{0}$ and $tr(\boldsymbol{A}_{1}^{\boldsymbol{Z}} \boldsymbol{Z})\ge{0}$, we can conclude
${P}^{*} (\boldsymbol{Z})\ge{P}^{*}-tr(\boldsymbol{A}_{1}^{\boldsymbol{Z}} \boldsymbol{Z})$
In addition from the above equations we have:
\begin{align}
&tr(\boldsymbol{A}_{1})^{\boldsymbol{Z}}-tr(\boldsymbol{A}_{1}^{\boldsymbol{Z}} \boldsymbol{Z})\le{tr(\boldsymbol{A}_{1}^{0})}\to{P}^{*} (\boldsymbol{Z})\le{P}^{*} \tag{37}
\end{align}
Therefore, we have:
\begin{align}
&{P}^{*}-tr(\boldsymbol{A}_{1}^{\boldsymbol{Z}} \boldsymbol{Z})\le{P}^{*} (\boldsymbol{Z})\le{P}^{*}\tag{38}
\end{align}
When $\boldsymbol{Z}$ is small $\small{P}^{*} (\boldsymbol{Z})$ gets close to its lower bound $({P}^{*}-tr(\boldsymbol{A}_{1} \boldsymbol{Z}))$. As a result, it is better that  $\boldsymbol{Z}$ meet these two criteria i.e. $tr(\boldsymbol{A}_{1}^{\boldsymbol{Z}}\boldsymbol{Z})\ge{0}$ and being a matrix with a small entries.

\section{Appendix E}
 According to the definition of nuclear norm, (20) can be the nuclear norm of an assumptive matrix ($\boldsymbol{R}_{k})$ which is stated as $\boldsymbol{R}_{k}=\boldsymbol{G}_{k}^{l}{J}_{k}$. Due to the fact that the SVD decomposition of $\boldsymbol{J}_{k}$ equals to ${\boldsymbol{\Lambda} }_{k}{\boldsymbol{\Pi} }_{k}\boldsymbol{\Psi} _{k}^{H}$, $\boldsymbol{G}_{k}^{l}$ should be equal to ${{\boldsymbol{\Lambda} }_{k}}\boldsymbol{\Phi} _{k}^{l}\boldsymbol{\Lambda} _{k}^{H}\normalsize$ in order to equality of (20) and (21a) holds. Therefore, we have: 
\begin{equation}
\|\boldsymbol{R}_{k}\|_{*}={{\left\| {{\boldsymbol{\Lambda} }_{k}}{{\boldsymbol{\Phi} }^{l}_{k}}\boldsymbol{\Lambda} _{k}^{H}{{\boldsymbol{\Lambda} }_{k}}{{\boldsymbol{\Pi} }_{k}}\boldsymbol{\Psi} _{k}^{H} \right\|}_{*}}={{\left\| {{\boldsymbol{\Lambda} }_{k}}{{\boldsymbol{\Omega} }_{k}}\boldsymbol{\Psi} _{k}^{H} \right\|}_{*}}\tag{39}\\\
\end{equation}
where ${{\boldsymbol{\Omega} }_{k}}$ is a diagonal matrix and equal to the product of ${{\boldsymbol{\Pi} }_{k}}$ and ${{\boldsymbol{\Phi} }^{l}_{k}}$. According to (39), $\|\boldsymbol{G}_{k}^{l}\boldsymbol{J}_{k}\|_{*}=\sum\limits_{i=1}^{d}{\frac{p\times({{\delta }_{i}}({\boldsymbol{J}_{k}}))}{{{({{{\delta} }_{i}}(\boldsymbol{J}_{k}^{l})+\xi )}^{1-p}}}}$.
\bibliographystyle{ieeetr}
\bibliography{References}
\end{document}